# The data paper as a socio-linguistic epistemic object: A content analysis on the rhetorical moves used in data paper abstracts

Kai Li

School of Information Resource Management, Renmin University of China, Beijing, China

59 Zhongguancun St, Haidian District, Beijing, China, 100872

+86 13161285125; kai.li@ruc.edu.cn

Chenyue Jiao

School of Information Sciences, University of Illinois Urbana-Champaign, Illinois, USA

501 E. Daniel St, Champaign, IL, USA, 61820

1-217-305-2995; cjiao4@illinois.edu

## Abstract

The data paper is an emerging academic genre that focuses on the description of research data objects. However, there is a lack of empirical knowledge about this rising genre in quantitative science studies, particularly from the perspective of its linguistic features. To fill this gap, this research aims to offer a first quantitative examination of which rhetorical moves—rhetorical units performing a coherent narrative function—are used in data paper abstracts, as well as how these moves are used. To this end, we developed a new classification scheme for rhetorical moves in data paper abstracts by expanding a well-received system that focuses on English-language research article abstracts. We used this expanded scheme to classify and analyze rhetorical moves used in two flagship data journals, *Scientific Data* and *Data in Brief*. We found that data papers exhibit a combination of IMRaD- and data-oriented moves and that the usage differences between the journals can be largely explained by journal policies concerning abstract and paper structure. This research offers a novel examination of how the data paper, a novel data-oriented knowledge representation, is composed, which greatly contributes to a deeper understanding of data and data publication in the scholarly communication system.

# 1 Introduction

The growing amount of research data in the 21st century brings significant changes to the position of data in the scientific system, how scientific research is defined, and how it should be conducted (Callaghan et al., 2012; Gitelman, 2013; Hey et al., 2009). In particular, these recent developments call for greater data transparency in the scholarly ecosystem, as summarized into the FAIR Principles of research data stewardship, i.e., research data should be findable, accessible, interoperable, and reusable (Wilkinson et al., 2016). A major solution to these new needs is the concept of *data publication*, which refers to the pipeline through which data are transformed into discrete, well-documented, publishable, and citable objects (Parsons & Fox, 2013). Under this framework, the practice of publishing data objects as academic papers, or *data papers*, has been embraced by researchers from a broad spectrum of knowledge domains. A data paper is a "scholarly publication of a searchable metadata document describing a particular online accessible dataset, or a group of datasets, published in accordance with the standard academic practices" (Chavan & Penev, 2011, p. 3).

Data papers have been proven to be useful in bridging the gaps between the research system and scientific data. First, published articles are highly effective in attracting citations to data objects, making it easier for data works to receive credit (Borgman, 2016; Jiao & Darch, 2020). Second, academic journals offer a heavily used peer-review pipeline. Despite the known limitations of the peer review system (Lee et al., 2013), publishing data in academic journals is the easiest solution for data to be peer-reviewed (Huang et al., 2013; Mayernik et al., 2015). These reasons support the growing popularity of data papers across various knowledge domains. For example, Candela and colleagues' survey (2015) identified seven *dedicated, or exclusive, data journals* (journals that only publish data papers) and over a hundred mixed data journals.

As an emerging academic genre, data papers are still far from visible in our research system. A chief reason for the invisibility is a lack of understanding of the textual composition of data papers as an academic genre and how this composition differs from that of regular research articles. On the one hand, data papers are only supposed to describe the collection, processing, and contents of their respective datasets, instead of presenting a study design and results (Callaghan et al., 2012). This definition should give data papers a distinct structure from research articles, for which the IMRaD (introduction, method, results, and discussion) format has been dominant in academic journals during the past several decades (Sollaci & Pereira, 2004). On the other hand, data papers inevitably share many textual similarities with research articles, as both genres are produced under comparable social norms and contexts of scientific knowledge production, such as the peer review system (Li et al., 2020).

This dilemma makes the textual features of data papers an important topic in quantitative science studies, because such features bear strong theoretical implications about how knowledge is produced and communicated from a socio-technical perspective (Bazerman, 1983; Small, 1982; Swales, 1990), and understanding them will facilitate a smoother integration of data papers into the research system. First, socio-linguistic decisions made by researchers could help us better understand the nature of data papers as a representation of scientific knowledge distinct from research articles, and to appreciate their functions in the scholarly communication system. Second, data paper composition practices should be more thoroughly considered in future open data policies, because data papers have become one of the most important means of publishing research data. However, no empirical research has yet investigated data papers from this critical perspective.

To fill this gap, the present study offers a quantitative examination of rhetorical moves used in abstracts of data papers. The rhetorical move is commonly defined as a "discoursal or rhetorical unit that performs a coherent communicative function in a written or spoken discourse" (Swales, 2004, p. 268). It has been used as a basic research instrument in various research communities focusing on academic writing, ranging from applied linguistics to quantitative science studies (Bazerman, 1988; Cross & Oppenheim, 2006; Swales, 1990; Thelwall, 2019). In this research, we examined the abstracts of data papers because, first of all, the abstract is commonly regarded as an accurate summary of the content of a scientific publication (Hall, 2012). Moreover, the abstract is also the section most frequently accessed by readers and bears a strong news value (Berkenkotter & Huckin, 2016). For these reasons, there is a large body of empirical works examining various linguistic attributes of research article abstracts, especially the move structure of the abstract (Hyland, 2000b; Lorés, 2004; Salager-Meyer, 1992; Stotesbury, 2003).

In this work, we analyzed what rhetorical moves are used, and how they are used, in the abstracts of 360 data papers selected from two flagship data journals, *Scientific Data* and *Data in Brief*. Using this paper sample, the following research questions are pursued in this study:

**RQ1: What rhetorical moves are used in abstracts of data papers?** This question aims to propose a classification scheme for the rhetorical moves used in data paper abstracts, focusing especially on those moves that are not shared by research articles. In this research, we expanded a highly-cited scheme for rhetorical moves in English-language research paper abstracts (Hyland, 2000b) by manually examining 50 randomly-selected data paper abstracts. The revised classification system was further validated during the manual classification process. This new system offers a solid methodological foundation for the present work and any future works on similar topics.

**RQ2: How are these moves distributed in the two data journals?** This question strives to understand the cross-journal differences in the use of rhetorical moves. Based on the modified classification system, we manually classified all sentences in our paper sample to understand the distribution of rhetorical moves in the space of data paper abstracts. Specifically, we compared how the identified moves were used differently by the journals and sought to explain their differences using the journal policies related to data paper abstract and paper structure. We also examined how the use of moves evolves over time on the journal level. This question deepens our existing knowledge about data papers as an academic genre by introducing a more dynamic perspective.

**RQ3: What is the order in which these moves are structured in the abstract?** This question examines the distribution of rhetorical moves within the sequence of abstract texts, with the purpose of understanding how rhetorical moves are used to construct data paper abstracts in a more granular manner. We examined how different rhetorical moves are used in different positions in the abstract, which sheds light on the deeper textual characteristics of different rhetorical moves involved in data papers.

# 2 Literature review

## 2.1 Data publication and data papers

With the significant growth of research data, the demand for data sharing has also been increasing, so that more data can be used to support knowledge production and scientific reproducibility (Borgman, 2015). Data publication is an important instrument for making data easier to access, understand, and reuse (Parsons & Fox, 2013), to eventually support more effective reuse of data. Data papers have been generally recognized as a new form of data publication and have been increasingly adopted in many research communities (Candela et al., 2015; Gorgolewski et al., 2013). A major distinction between data papers and research article is their focus: while both types of publications are normally peer-reviewed, data papers are designed to describe data objects (Carlson & Oda, 2018), a major divergence from the IMRaD paper structure (Sollaci & Pereira, 2004).

The unique structure of data papers can greatly contribute to data sharing and reuse. First, effective data sharing requires not only the availability of the data object per se, but also sufficient metadata to justify the reusability of the data (Zimmerman, 2007). Data papers thus function as an important incentivizing instrument for researchers to undertake the preparation and sharing of data and its accompanying metadata (Gorgolewski et al., 2013). Second, despite the different paper structures across journals, most require authors to detail methods used to produce the data object and the repository of the dataset (Kim, 2020). Detailed information about how the data object was collected, processed, and shared can support the discovery and evaluation of data papers by other researchers (Faniel et al., 2019; Wang et al., 2021).

During the past few years, data papers have become an increasingly popular type of scientific publication. Candela et al. (2015) identified over 100 journals that accept data papers, and there is evidence that the number of data papers has continued to increase since then (El-Tawil & Agrawal, 2019). Empirical evidence has also shown that data papers are frequently cited, although the citation pattern normally takes on a highly long-tailed distribution (Kotti & Spinellis, 2019) and the papers may be cited for reasons other than reuse of the data (Jiao & Darch, 2020). Much more empirical investigation, using quantitative and bibliographic data, is needed before we can comprehensively understand the data paper's roles and positions in the scholarly communication system, especially as distinct from regular research articles. The present paper, focusing on the rhetorical moves of data papers, is a step towards this goal.

## 2.2 Research paper abstract as an academic genre

The abstract is an indispensable component of research articles and is required by most journals as part of an article submission (Martín, 2003). Hyland (2015) observes that "the abstract is generally the readers' first encounter with a text, and is often the point at which they decide whether to continue and give the accompanying article further attention or to ignore it" (p. 63). Underlying this phenomenon is the function of abstracts as a somewhat accurate summary of research articles (Hall, 2012; Salager-Meyer, 1990). This "news value" has been further highlighted in recent years as the pace of academic publication accelerates (Landhuis, 2016).

Given their centrality in the genre of research articles, research article abstracts have been heavily studied in the applied linguistics community, particularly from the perspective of rhetorical moves. For example, dos Santos (1996) and Hyland (2000b) have each proposed a five-step classification system for rhetorical moves used in English-language research article abstracts. The systems are similar in that they cover the most basic functional units in the IMRaD structure, but Hyland's system is regarded as superior because it was built on a much larger sample size from a broader selection of knowledge domains (Suntara & Usaha, 2013).

Apart from these two classification systems, a large number of empirical works have examined how rhetorical moves are used to construct research article abstracts in different knowledge domains and/or countries (Dahl, 2004; Gillaerts & van de Velde, 2010; Hyland, 2000a; Lorés, 2004; Saeeaw & Tangkiengsirisin, 2014; Samraj, 2002; van Bonn & Swales, 2007). Despite their different scopes, most of these studies (1) have the common aim of supporting pedagogical activities concerned with English-language research article writing and (2) are based on relatively small paper samples and the manual coding method.

Moreover, this line of research is situated in a socio-linguistic approach to scientific documents that can have meaningful conversations with quantitative science studies. Earlier works have shown how textual features, when combined with other quantitative methods (especially the citation analysis method), serve as important indicators of the research community of publications and their roles in scholarly communication (Swales, 1986; White, 2004). We believe this approach will likewise contribute to a deeper understanding of the emerging and heretofore rarely studied genre of data papers. This paper is the first step towards this novel research direction.

## 3 Method

### 3.1 Data collection

In this study, we analyzed data papers included in *Scientific Data* and *Data in Brief*, two domain-independent flagship data journals. The two journals were selected because, firstly, they are the two leading exclusively data journals, based on journal impact factor, number of publications, and presence in empirical studies (Kim, 2020; Walters, 2020). For this reason, we believe these journals offer a representative sample of data publications. Secondly, both journals were founded in 2014, which provides a meaningful time frame to conduct temporal analyses. We collected all metadata records in these journals from Scopus on November 15, 2020.

While both journals were categorized as exclusively data journals by Stuart (2017), they contain document types beyond data papers. As an illustration, Scopus classified most publications in *Data in Brief* as *Data Papers*, but almost 75% of *Scientific Data* publications as *Articles*. We validated the Scopus classification with the information on the journal websites. We found that the "data descriptors," as they are classified by *Scientific Data*, were categorized as either *Data Papers* or *Articles* by Scopus, whereas most data papers in *Data in Brief* were classified as such in Scopus. Based on this information, we removed all publications that are NOT classified as data papers by the journal websites, including both research articles and other publication types (most notably, errata and editorial). Our final sample includes 7,712 data papers from both journals, with 6,335 from *Data in Brief* and 1,377 from *Scientific Data*.

From these journals, we selected only data papers published between 2015 and 2020, as both journals were founded in 2014 and there are insufficient papers to analyze in that year. To enable a more meaningful comparison between the journals, we used a stratified sampling approach in which we randomly selected 30 publications per journal per year. There are 360 data papers in our final sample for manual classification.

For all selected publications, we used the NLTK Python package (Loper & Bird, 2002), a popular natural language processing library used on scientific corpora, to parse abstracts into sentences. After the automatic processing, we examined the results and manually fixed any mistakenly parsed sentences. A total of 2,182 sentences were parsed and analyzed in the following steps.

## 3.2 Manual classification of rhetorical moves

This research aims to identify rhetorical moves used in data paper abstracts on the sentence level. For this purpose, we expanded the classification scheme proposed by Hyland (2000b) for rhetorical moves in English-language research article abstracts, which is composed of *Introduction*, *Purpose*, *Method*, *Product*, and *Conclusion*.

To align this scheme with data papers, two coders manually coded 50 randomly-selected abstracts (from the original 7,712 papers) and independently identified new categories not covered by the scheme. In the end, we added four new rhetorical moves found in our paper sample. Our modified scheme is shown in Table 1, where we highlight the four added moves and state how they are defined in this project. Moreover, we changed *Product* to *Results* in our scheme while retaining the original definition proposed by Hyland. This scheme was tested during our manual classification, but we did not find any other category to add.

**Table 1: Classification and definition of rhetorical moves in data paper abstracts**

| Move | Definition |
|---|---|
| Introduction | Context of the paper |
| Purpose | Purpose or intention of the paper/research |
| Method | Research design, procedure, assumptions, approach of the study |
| Results | Main findings or results |
| Conclusion | Interpretations of the results beyond the scope of the paper |
| **Data Description** | Description of the data object that is the topic of the paper |
| **Data Uses** | Proposed uses or implications of the data object |
| **Data Accessibility** | How to get access to the data object |
| **Related Research Article** | The research article to which the data object is connected |

The majority of the new moves focus on research data, in contrast to the traditional IMRaD-oriented moves. These two orientations support the data publication genre, which follows the basic rhetoric of research articles but is dedicated to the description, discovery, and reuse of data objects. We also identified the move of Related Research Article, which serves to link a data paper to another research article where the dataset was initially used. Examples of these moves from our paper sample are offered in Table 2.

Table 2: Examples of rhetorical moves as used in our paper sample

| Move | Example |
|---|---|
| Introduction | "*Aspergillus* sp. plays an important role in lignocellulosic biomass recycling." (Adav et al., 2015) |
| Purpose | "This data article describes a detailed synthetic strategy and experimental data for the synthesis of brominated fulvene chromophores..." (Budy et al., 2018) |
| Method | "Firstly, we establish a robust approach to extracting shorelines from vertical aerial photography, validated against LiDAR (Light Detection and Ranging) and coastal topography surveys." (Pollard et al., 2019) |
| Results | "We found that the depletion of Jumonji domain-containing protein 6 (JMJD6) […] slowed cell proliferation of mouse NIH3T3 fibroblasts." (Hu & Imbalzano, 2016) |
| Conclusion | "As coastal systems are monitored at greater spatial resolution and temporal frequency there is an unprecedented opportunity to determine how and why coastal systems have changed in the past with a view to informing future forecasting." (Pollard et al., 2019) |
| Data Description | "The data presented here represents the detailed comparative abundances of diverse groups of biomass hydrolyzing enzymes […] and their post translational modification like deamidation." (Adav et al., 2015) |
| Data Uses | "The Australian Phytoplankton Database […] allows analysis of ecological indicators of climate change and eutrophication (e.g., changes in distribution; diatom:dinoflagellate ratios)." (Davies et al., 2016) |
| Data Accessibility | "We have lodged this dataset with the Australian Ocean Data Network (http://portal.aodn.org.au/) allowing public access." (Davies et al., 2016) |
| Related Research Article | "Here, we provide data related to the research article entitled 'Quantitative proteomics study of *Aspergillus fumigatus* secretome revealed deamidation of secretory enzymes' by Adav et al. (J. Proteomics (2015) [1])." (Adav et al., 2015) |

Two coders independently classified all sentences using the modified classification scheme. Our coding also allows the co-existence of multiple moves in one sentence. In the final coding, 77 sentences contain two moves. For moves in these sentences, we used fractional counting to measure the frequencies of moves on the sentence level, counting each move as 0.5 instead of one sentence. The intercoder reliability is 0.706, which indicates a good agreement (Landis & Koch, 1977). All differences between the coders were resolved before data analysis.

## 3.3 Abstract-related journal policies

To understand how the rhetorical moves reflect the policies of the two data journals, we also collected policy statements related to data paper abstracts from both journals, focusing on the following two broad areas:

**Abstract content and structure**: This area concerns policies that directly regulate the content and structure of data paper abstracts. As shown in Table 3, the journals have very similar policies, except that *Scientific Data* requires that no reference be given in the abstract.

Table 3: Journal policies concerning content and structure of data paper abstracts

| Policy Area | *Scientific Data* | *Data in Brief* |
|---|---|---|
| Abstract content and structure | - They should succinctly describe the study, the assay(s) performed, the resulting data and their reuse potential.<br>- It should not make any claims regarding new scientific findings.<br>- No references are allowed in this section. [1] | - Concisely describes the data, its collection process, analysis and reuse potential.<br>- Do not: provide conclusions, results, or mention the word 'study'. [2] |

**Paper structure**: We also collected paper structure specifications to understand how the paper structure influences the uses of rhetorical moves in paper abstracts. This takes into account the fact that most traditional rhetorical moves correspond to major IMRaD paper sections. As summarized in Table 4, the two data journals have distinct mandated paper structures. It should be noted that these structures are consistently used in their respective journals based on our manual examination and previous evidence (Li & Chen, 2018).

Table 4: Journal policies concerning the structure of data papers

| Journal | Paper Section | Definition |
|---|---|---|
| *Scientific Data*[3] | Background & Summary | This section should provide an overview of the study that generated the data, as well as outlining the potential reuse value of the data. |
| | Methods | The Methods section in Data Descriptors should describe any steps or procedures used in producing the data. |
| | Data Records | This section should be used to explain each data record associated with this work, including the repository where this information is stored, and to provide an overview of the |

---

[1] https://www.nature.com/sdata/publish/submission-guidelines
[2] https://www.elsevier.com/__data/promis_misc/DIB%20Article%20Template%203.2.1.docx

[3] The required paper structure as well as the specific requirements of *Scientific Data* are taken from: https://www.nature.com/sdata/publish/submission-guidelines.

|  |  | data files and their formats. |
|  | Technical Validation | This section should present any experiments or analyses that are needed to support the technical quality of the dataset. |
|  | Usage Notes | 'Usage Notes' is an optional section that can be used to provide information that may assist other researchers who reuse your data. |
| *Data in Brief*[4] | Value of Data | - Why are your data useful or important to the research community?<br>- Who can use/benefit from the data?<br>- How can the data be (re)used for further insights or analysis? |
|  | Data Description | Briefly, describe and refer to each data file included in the article and supplementary files or in the repository. |
|  | Experimental Design, Materials, and Methods | Provide a detailed description of how the data were acquired and analyzed. Aid data reproduction and reuse by offering a more comprehensive description than any related article or previous articles. |

### 3.4 Descriptive analysis: length and complexity of data paper abstracts

Figure 1 summarizes the frequency of papers with specific numbers of rhetorical moves in the paper abstract. The graph shows a normal distribution, where over 80% of articles have three to five rhetorical moves. This pattern is relatively similar across the two journals: the mean number of moves per abstract is 3.85 in *Scientific Data* and 3.57 in *Data in Brief*. This difference mirrors the difference in abstract length between these journals (mean of 6.38 sentences in *Scientific Data* and 5.74 in *Data in Brief*).

**Figure 1: Distribution of data papers by number of moves in the abstract (*n* = 360)**

---

[4] The required paper structure of *Data in Brief* is taken from:
https://www.elsevier.com/__data/promis_misc/DIB%20Article%20Template%203.2.1.docx. Specifications of the paper sections are taken from:
https://www.elsevier.com/__data/promis_misc/Data%20in%20Brief%20Article%20at%20a%20Glance%20illustration.pdf.

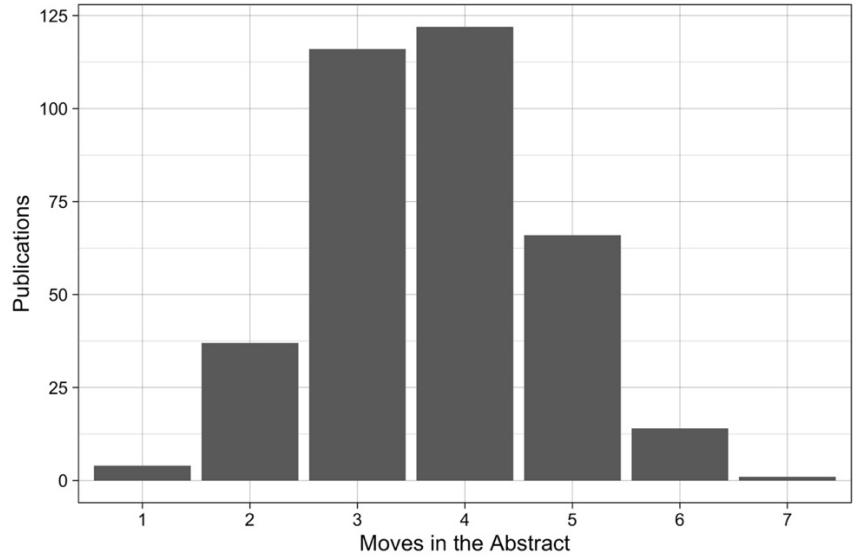

## 4 Results

### 4.1 How often are the rhetorical moves used in data paper abstracts?

Table 5 illustrates how the rhetorical moves are distributed in our sample. Introduction, Method, Data Description, and Data Uses are seen to be the most frequently used moves on both the sentence and paper levels. This group of moves covers both IMRaD- and data-oriented functions. A notable finding, contrasting with the journal policies, is that the category of Results is frequently used in abstracts: more than 25% of articles have at least one Results sentence. Following our previous work (Li et al., 2020), this finding challenges the boundaries between research articles and data papers in terms of paper content.

Table 5: Distribution of rhetorical moves in our sample

| Rhetorical Moves | Sentences ($n$ = 2,182) | Papers ($n$ = 360) |
|---|---|---|
| Introduction | 514 | 217 |
| Method | 527 | 249 |
| Data Description | 357 | 233 |
| Data Uses | 227.5 | 188 |
| Results | 185.5 | 97 |
| Purpose | 149 | 139 |
| Related Research Article | 93.5 | 93 |
| Conclusion | 67.5 | 52 |
| Data Availability | 61 | 67 |

Based on Table 5, we compared how the IMRaD moves are distributed in data papers versus research articles. As the control group, we selected two recent works, Amnuai (2019) and

Farzannia & Farnia (2017), both of which examined rhetorical moves used in English-language journal research articles based on Hyland's classification system. The two works focus on accounting and mining engineering, respectively. We acknowledge the difficulty of finding empirical works that can directly match the scope of the journals analyzed in the present work, given the multi-disciplinary nature of both data journals. As such, this comparison is only a preliminary step towards a more comprehensive understanding of the differences between data papers and research articles.

As shown in Table 6, the two distributions based on research articles are highly similar to each other; Purpose, Method, and Results are used in more than 80% of articles in each sample. By comparison, our sample of data papers does not have any dominant categories, partly because of the larger number of categories used in data papers. Purpose, Results, and Conclusion categories are all used much less extensively in data papers than in the two baseline samples; in the latter two cases, this may be explained by the fact that these categories are forbidden by journal policies. On the other side of the story, Introduction is used similarly across corpora, which may indicate that it is one of the most basic building blocks of narratives in both genres.

Table 6: Comparison of IMRaD move use in three paper samples

| Rhetorical Moves | Data Papers ($n = 360$) | Amnuai (2019) ($n = 30$) | Farzannia & Farnia (2017) ($n = 30$) |
|---|---|---|---|
| Introduction | 60.2% | 60% | 50% |
| Purpose | 38.6% | 90% | 86.7% |
| Method | 69.1% | 80% | 90% |
| Results | 26.9% | 90% | 93.3% |
| Conclusion | 14.4% | 46.7% | 63.3% |

Table 7 summarizes how the moves are used differently by the two journals. There are two notable differences: first, Related Research Article is only used in *Data in Brief*. This can be explained by the fact that *Data in Brief* asks authors to include a citation to the research article, when the data paper is submitted via another Elsevier journal (co-submission)[5]. However, references are specifically forbidden in *Scientific Data*.

Table 7: Summary of rhetorical move use in the two data journals

| | *Data in Brief* | | *Scientific Data* | |
|---|---|---|---|---|
| Moves | Sentences ($n = 1033$) | Papers ($n = 180$) | Sentences ($n = 1149$) | Papers ($n = 180$) |
| Introduction | 169 | 76 | 345 | 141 |
| Method | 303.5 | 130 | 223.5 | 119 |
| Data | 142.5 | 101 | 214.5 | 132 |

---

[5] https://www.elsevier.com/__data/promis_misc/DIB%20Article%20Template%203.2.1.docx

| Description | | | | |
|---|---|---|---|---|
| Data Uses | 61 | 53 | 166.5 | 135 |
| Results | 115.5 | 52 | 70 | 45 |
| Purpose | 79.5 | 77 | 69.5 | 62 |
| Related Research Article | 93.5 | 93 | 0 | 0 |
| Conclusion | 39.5 | 29 | 28 | 23 |
| Data Availability | 29 | 31 | 32 | 36 |

Second, both Introduction and Data Uses are adopted much more heavily in *Scientific Data*, which echoes the inclusion of corresponding paper sections in *Scientific Data*'s paper structure. Introduction directly corresponds to the Background & Summary section; Data Uses, to the Usage Notes section. Moves covered by both structures, especially Methods and Data Description, seem to be used similarly across journals. Even though it is difficult to find a direct match between every rhetorical move and paper section, there seems to be a general relationship between the availability of a paper section and how its corresponding move is used in the abstract.

Figure 2 shows the temporal trend of how moves are used on the journal level. We found that most of the moves are relatively consistently used over time by the two journals, despite yearly fluctuations, indicating a relatively stable landscape in the paper sample we examined.

**Figure 2: Use of rhetorical moves in the two journals over time**

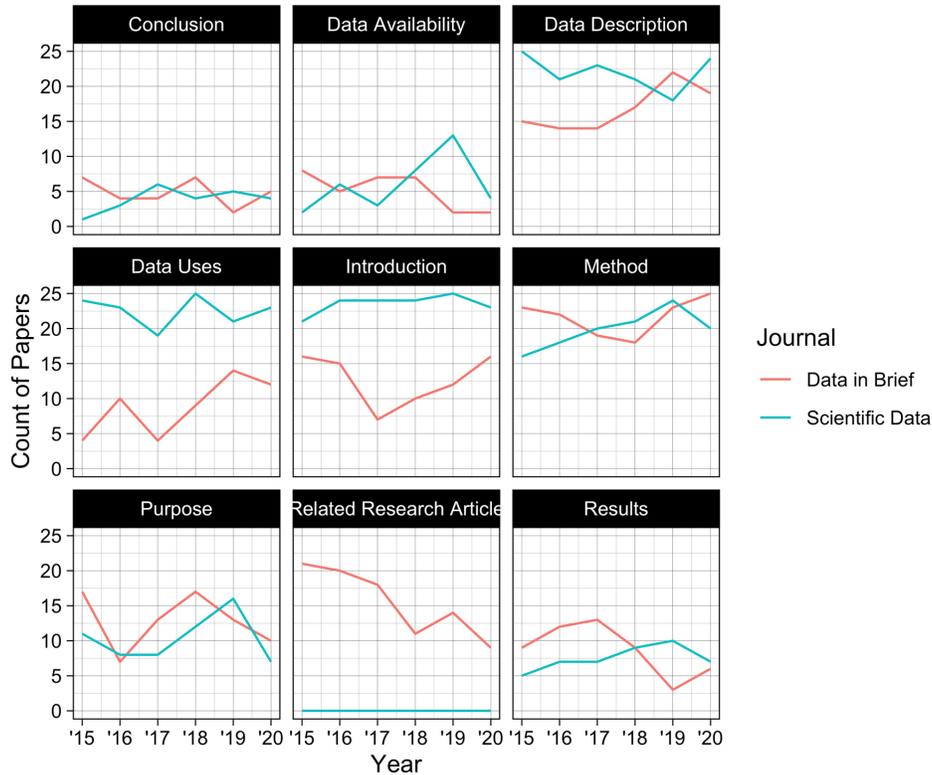

## 4.2 Order of rhetorical moves in abstracts

To understand the order of rhetorical moves in data paper abstracts, we measured which moves are used as the first and last moves in the abstract. Table 8 shows the top five moves used as the first move in the abstract in our overall sample (left panel), *Data in Brief* (middle panel), and *Scientific Data* (right panel). The results show that Introduction is the opening move in more than 50% of abstracts, followed by Data Description and Related Research Article. However, this pattern is vastly different across the two journals: in *Scientific Data*, nearly 75% of articles use Introduction as the first move, while this number is lower than 40% in *Data in Brief*. Aligned with results in the previous section, this difference can be explained by the fact that *Scientific Data* has an Introduction section in the beginning of its paper structure. In *Data in Brief*, however, Related Research Article, Data Description, and Purpose can frequently be used to start the abstract despite their quite different rhetorical functions. This result suggests a higher level of standardization in data paper abstracts in *Scientific Data*. Moreover, it is worth noting that the categories of Results, Data Availability, and Conclusion are never used as the first move in the abstract.

Table 8: Most frequent first moves in data paper abstracts from the full sample and individual journals

| Full Sample | | *Data in Brief* | | *Scientific Data* | |
| --- | --- | --- | --- | --- | --- |
| Move | Percentage of Articles | Move | Percentage of Articles | Move | Percentage of Articles |
| Introduction | 55.6% | Introduction | 36.7% | Introduction | 74.4% |

| Data Description | 16.9% | Related Research Article | 21.1% | Data Description | 13.9% |
| Related Research Article | 10.6% | Data Description | 20% | Method | 4.4% |
| Purpose | 8.3% | Purpose | 13.3% | Purpose | 3.3% |
| Method | 4.7% | Method | 5% | Data Uses | 1.1% |

Table 9 shows the distribution of last moves, following the structure of Table 8. The results show that the last move is much less consistent than the first; however, as in Table 8, *Scientific Data* is more standardized than *Data in Brief*, as illustrated by the fact that Data Uses appears as the last move in more than 50% of its publications, whereas no single move closes the abstract of more than 20% of *Data in Brief* publications. The fact that Data Uses is the most frequently used last move in *Scientific Data* may be partly explained by the fact that Usage Notes is the last section in its paper structure.

**Table 9: Most frequently last moves in data paper abstracts from the full sample and individual journals**

| Full Sample | | *Data in Brief* | | *Scientific Data* | |
| --- | --- | --- | --- | --- | --- |
| Move | Percentage of Articles | Move | Percentage of Articles | Move | Percentage of Articles |
| Data Uses | 37.5% | Data Uses | 19.4% | Data Uses | 55.6% |
| Data Description | 9.4% | Related Research Article | 17.8% | Conclusion | 9.4% |
| Conclusion | 9.2% | Results | 12.2% | Data Description | 8.3% |
| Data Availability | 9.2% | Method | 11.7% | Data Availability | 7.8% |
| Related Research Article | 8.9% | Data Description | 10.6% | Method | 5% |

Figure 3 shows the distribution of moves in each move sequence when move complexity is between three and five, the three most dominant levels based on Figure 1. The color of each box represents the percentage of all moves in the position that belong to a specific category. For example, in all three scenarios, Introduction is the most likely to be used as the first move in the abstract. Overall, the graph shows a strong similarity across the three scenarios, which suggests that even though new moves may be added when the abstract becomes longer, the basic relationship between rhetorical moves and data paper abstracts remains consistent.

**Figure 3: Likelihood of rhetorical move use by position in abstracts with three to five moves**

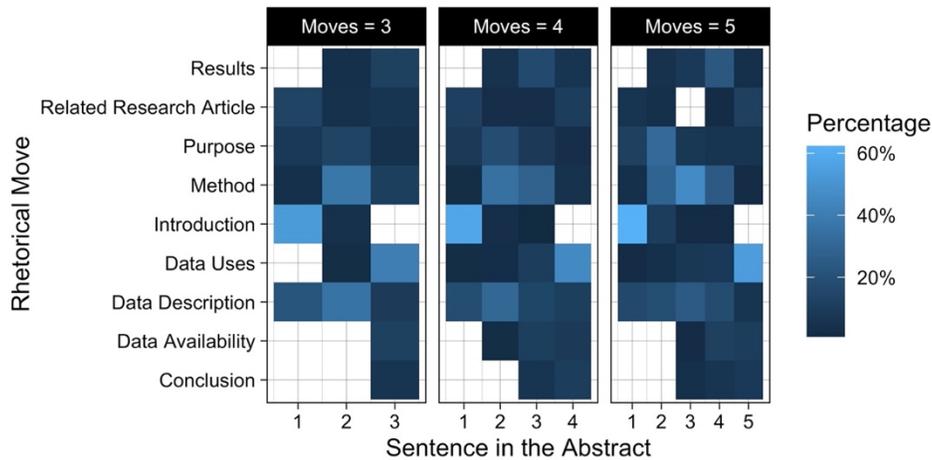

## 5 Discussion

### 5.1 Rhetorical moves used in data papers

The present research strives to understand what rhetorical moves are used in the emerging genre of data paper abstracts. By using the manual coding method, we found that data paper abstracts involve both traditional IMRaD rhetorical moves and new data-focused moves. These two general categories support the two-sided identity of the data paper: that is, it is both a scientific publication following traditional writing norms and a descriptor of data objects. These two aspects of identity embodied in the genre of data papers are not easily distinguishable from each other based on our results as well as earlier evidence (Li et al., 2020).

Our results show that traditional rhetorical moves, as summarized by Hyland (2000b), still have a strong presence in data paper abstracts. For example, Introduction and Methods are the two most frequently used moves in our sample, each appearing in over 60% of examined articles. Moreover, even though both journals specifically required that results not be discussed in data papers, we find that the Results move is used in about 25% of data papers in the sample. One potential explanation of this ostensible difference is that results may have different meanings in these genres, which warrants a future study. Overall, these results indicate that the traditional building blocks of narratives in research articles remain useful in constructing the new genre of data papers, even though they are no longer the sole focus of the new genre of data papers.

More interestingly, data papers introduce new rhetorical moves that are strongly related to the goals of promoting data object sharing and reuse (Jiao & Darch, 2020), including Data Description, Data Uses, Data Availability, and to a lesser extent, Related Research Article. The first three moves serve to offer more granular descriptions of the data objects that are the focus of data papers. Data Description and Data Uses are each used in more than 50% of articles, showing the popularity of these moves. Moreover, both Data Description and Data Uses have been increasingly used in our sample. Generally, these data-related moves have become important components in the narratives of data papers, where they are combined with traditional blocks in the IMRaD structure.

## 5.2 How these moves are used across journals and over time

Based on our modified classification scheme, this paper quantitatively examines how the rhetorical moves are used by the two selected journals, to gain a dynamic understanding of the genre of data papers. Our results testify that there is general consistency in how rhetorical moves are used in data publications. Most of the move categories are used by both journals and many of them are used similarly across journals, in terms of both frequency and order. This indicates that the data paper can be regarded as a distinct and somewhat stabilized academic genre in terms of its linguistic features.

Despite this general consistency, however, there are some important differences between the journals, which sheds important light on the diversity of genre of data papers. Moreover, we found that many of these differences can be at least partly explained by journal policies about how the abstract should be constructed or the overall paper structure.

One of the starkest differences lies on the fact that the Related Research Article move is only used in *Data in Brief*, which can be explained by the fact that this journal allows a reference to another research article that is co-submitted with the data paper, whereas *Scientific Data* specifically forbids the use of references in paper abstracts. Moreover, the much higher usage of Introduction and Data Uses moves in *Scientific Data* mirrors the use of paper sections in this journal with similar rhetorical functions, which supports the abstract's role as a somewhat accurate summary of the paper content as reported in applied linguistics (Hall, 2012).

On a higher level, these differences indicate the diversity of practices in using the emerging genre of data papers to describe data objects in the current socio-technical environment. One consequence of this diversity is that different journal norms or policies may not have the same capacity to support the discovery and reuse of data objects, which is an important topic to study in future works. Moreover, it can be expected that more conversations will be needed in the future to develop more fine-grained policies about what data papers should describe, as this genre becomes more popular and undertakes greater responsibility in scholarly communication.

## 6 Conclusion

This paper offers a quantitative study of the rhetorical moves used in data paper abstracts, by analyzing 360 data papers acquired from *Scientific Data* and *Data in Brief*, two flagship data journals. We developed a classification scheme for rhetorical moves used in data papers by expanding a similar scheme focusing on research paper abstracts. Using the scheme as the basis of manual classification, we analyzed how these moves are used between the two journals. We found that IMRaD-oriented moves and data-focused moves are mixed in data paper abstracts, supporting a two-sided and intertwined identity of this emerging genre. Moreover, in comparing the journals, we found that journal policies are a major factor underlying the different usage of these moves, which may have long-term implications for the communication of data-focused knowledge and outputs in the research system.

This research addresses the urgent need to understand data papers as an emerging academic genre in the scholarly communication system, particularly from a socio-linguistic perspective, because of the increasing popularity and unique linguistic features of this genre. This work illustrates the diversity and uncertainty of practices in the early years of a new data-centric academic genre and shows how socio-technical factors, especially journal policies, influence these practices. As a result, this work sheds light on how research data is involved in new modes of knowledge production and highlights important issues in the identity of the data publication. Our results will also serve as an important baseline dataset for future works on the rhetorical structure of data papers, especially those focusing on the disciplinarity of data papers and the differences between data papers and research articles.

Despite its timely contributions, this paper is only the first step towards the goal of understanding data publication as rising academic practices and outputs, and it has a few limitations that should be addressed in future works. First, we only analyzed two domain-independent data journals in this work and did not consider how the disciplinarity of data papers influences their textual features. We acknowledge that academic writing practices are deeply situated in the disciplinary contexts, as shown in earlier empirical evidence (Saeeaw & Tangkiengsirisin, 2014; Suntara & Usaha, 2013). Given the limitations of journal-based classification systems implemented in the Web of Science or Scopus (Shu et al., 2019), we believe that this issue will need to be addressed on the infrastructure level, by developing a more effective approach to classifying data papers by research domain. Second, this study offers only a very preliminary comparison between data papers and research articles; future work in this vein will greatly contribute to a more accurate distinction between data papers and research articles in scientific databases, especially in mixed data journals (i.e., journals accepting both genres discussed above). We plan to address both of these limitations in the next step of our project to better situate data papers in our existing scholarly communication system.